\documentclass[conference]{IEEEtran}
\IEEEoverridecommandlockouts
\usepackage{cite}
\usepackage{amsmath,amssymb,amsfonts}
\usepackage{algorithmic}
\usepackage{graphicx}
\usepackage{textcomp}
\usepackage{xcolor}
\usepackage{booktabs}
\usepackage{multirow}
\usepackage{array}
\usepackage{subfigure}
\usepackage{url}
\usepackage{algorithm}
\usepackage{algorithmic}
\usepackage{hyperref}

\def\BibTeX{{\rm B\kern-.05em{\sc i\kern-.025em b}\kern-.08em
    T\kern-.1667em\lower.7ex\hbox{E}\kern-.125emX}}

\begin{document}

\title{Multi-Agent Medical Decision Consensus Matrix System: An Intelligent Collaborative Framework for Oncology MDT Consultations}

\author{\IEEEauthorblockN{Xudong Han}
\IEEEauthorblockA{
\textit{University of Sussex}\\
xh218@sussex.ac.uk}
\and
\IEEEauthorblockN{Xianglun Gao}
\IEEEauthorblockA{
\textit{Columbia University}\\
tpjcw0309@gmail.com}
\and
\IEEEauthorblockN{Xiaoyi Qu}
\IEEEauthorblockA{
\textit{Lehigh University}\\
xiq322@lehigh.edu}
\and
\IEEEauthorblockN{Zhenyu Yu}
\IEEEauthorblockA{
\textit{University of Malaya}\\
yuzhenyuyxl@foxmail.com}
}

\maketitle

\begin{figure*}
    \centering
    \includegraphics[width=0.8\linewidth]{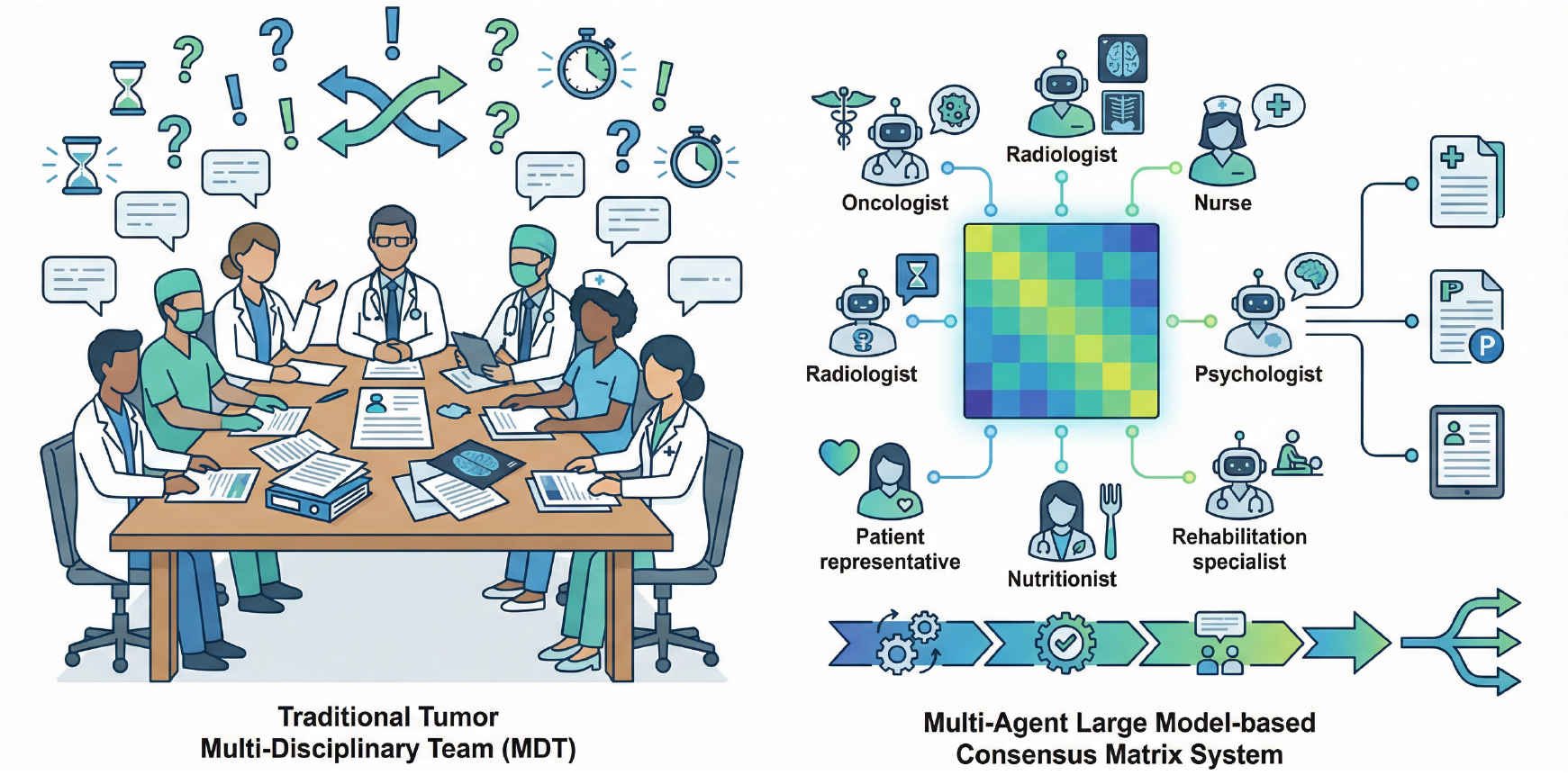}
    \caption{Motivation. Traditional tumor multidisciplinary team (MDT) meetings rely on unstructured discussion among human experts, often lacking quantitative consensus measurement and traceable evidence for treatment decisions. Our framework replaces each MDT role with a specialized large language model agent, aggregates their structured preferences into a mathematically grounded consensus matrix, and uses reinforcement learning together with guideline- and literature-based evidence retrieval to produce oncology treatment recommendations with quantified agreement and full evidence traceability.}
    \label{fig:motivation}
\end{figure*}

\begin{abstract}
Multidisciplinary team (MDT) consultations are the gold standard for cancer care decision-making, yet current practice lacks structured mechanisms for quantifying consensus and ensuring decision traceability. We introduce a Multi-Agent Medical Decision Consensus Matrix System that deploys seven specialised large language model agents, including an oncologist, a radiologist, a nurse, a psychologist, a patient advocate, a nutritionist and a rehabilitation therapist, to simulate realistic MDT workflows. The framework incorporates a mathematically grounded consensus matrix that uses Kendall’s coefficient of concordance to objectively assess agreement. To further enhance treatment recommendation quality and consensus efficiency, the system integrates reinforcement learning methods, including Q-Learning, PPO, and DQN. Evaluation across five medical benchmarks (MedQA, PubMedQA, DDXPlus, MedBullets, SymCat) shows substantial gains over existing approaches, achieving an average accuracy of 87.5\% compared with 83.8\% for the strongest baseline, a consensus achievement rate of 89.3\%, and a mean Kendall’s $W$ of 0.823. Expert reviewers rated the clinical appropriateness of system outputs at 8.9/10. The system guarantees full evidence traceability through mandatory citations of clinical guidelines and peer-reviewed literature following GRADE principles. This work advances medical AI by providing structured consensus measurement, role-specialised multi-agent collaboration, and evidence-based explainability to improve the quality and efficiency of clinical decision-making.
\end{abstract}


\section{Introduction}

Cancer care remains one of the most intricate domains in modern medicine, requiring the coordinated integration of expertise across multiple specialties. Multidisciplinary Team (MDT) consultations have consequently become a central paradigm for oncological treatment planning, bringing together oncologists, radiologists, pathologists, nurses, and other professionals to formulate joint therapeutic strategies~\cite{prades2015multidisciplinary, wu2024novel}. A series of clinical studies have shown that MDT-driven pathways can improve survival by 15--25\%~\cite{fennell2010multidisciplinary, kesson2012effects, xin2024v, xin2024mmap, xin2023self, yang2025wcdt}, reduce treatment delays by an average of 8.3 days, and enhance care coordination through the adoption of standardised protocols~\cite{pillay2016multidisciplinary, lin2025hybridfuzzingllmguidedinput}.

Despite these benefits, everyday MDT practice still faces substantial operational and methodological challenges. In many hospitals, tumour board meetings rely on loosely structured discussion, without systematic mechanisms for synthesising diverse expert opinions, quantifying the strength of consensus, or documenting the reasoning that underpins collective decisions~\cite{lamb2013improving, xin2024parameter, xin2025lumina}. This lack of formal structure has tangible consequences: recommendations may be inconsistent in roughly a quarter of cases, discussion time per patient can be lengthy, and the traceability of decision-making is often poor, which in turn complicates quality assurance and retrospective review~\cite{huncovsky2024cognitive, yu2025cotextor, he2025ge}.

In parallel, recent breakthroughs in artificial intelligence, particularly in Large Language Models (LLMs), have demonstrated strong capabilities in medical reasoning and clinical decision support~\cite{niu2024textmultimodalityexploringevolution, yan2025largelanguagemodelbenchmarks, wang2025twin}. For instance, GPT-4~\cite{team2024gemma, gao2025free, xin2024vmt, wu2020dynamic} has achieved an accuracy of 86.4\% on USMLE Step 1 examinations~\cite{nori2023capabilities}, while Med-PaLM 2 has exhibited expert-level performance on medical licensing benchmarks~\cite{singhal2023towards, wang2023intelligent, zhou2025reagent}. Beyond single-model setups, multi-agent systems have shown considerable promise in modelling collaborative decision-making processes, enabling the simulation of expert interaction, critique, and consensus formation across a range of domains~\cite{nweke2025multi,yu2025visualizing}.

However, contemporary multi-agent medical systems still exhibit several critical limitations. First, many frameworks provide only rudimentary voting-based schemes for combining agent outputs, lacking structured metrics for characterising consensus beyond simple tallies~\cite{nweke2025multi,yu2025forgetme}. Second, role specialisation is often coarse and fails to capture the nuanced differences in expertise, responsibility, and perspective that exist in real-world medical teams. Third, explainability and evidence traceability remain limited, as systems frequently offer free-form rationales with weak links to clinical guidelines or primary literature, which constrains clinical acceptance. Finally, the absence of adaptive learning mechanisms means that performance is not systematically improved over time~\cite{wu2024tutorial, cao2025tv, xin2025resurrect, lin2025llmdrivenadaptivesourcesinkidentification, liang2025sage}. Existing approaches such as MDAgents~\cite{kim2024mdagents, tian2025centermambasamcenterprioritizedscanningtemporal} and TeamMedAgents~\cite{mishra2025teammedagents, cao2025cofi, wu2024augmented} mark important progress, but they still rely primarily on majority voting or basic opinion aggregation and pay limited attention to the quality and coherence of consensus formation.

To address these deficiencies, we propose a novel \textbf{Multi-Agent Medical Decision Consensus Matrix System}. The framework introduces a structured matrix representation that encodes expert opinions as quantified treatment preferences, clinical reasoning, confidence scores, and documented concerns, thereby enabling objective consensus measurement using Kendall's coordination coefficient. Our architecture deploys seven distinct medical role agents with tailored professional characteristics and interaction protocols, mirroring the dynamics of real-world MDTs. We model the consensus formation process as a Markov Decision Process and apply advanced Reinforcement Learning methods such as Q-Learning, PPO and DQN to enhance both recommendation quality and consensus efficiency~\cite{wu2022adaptive}. Each agent's opinion is supported by a traceable evidence chain that cites clinical guidelines including NCCN and ESMO as well as peer-reviewed studies from PubMed, ensuring transparency and clinical credibility (see Figure~\ref{fig:motivation}). 

Our main \textbf{contributions} are:
\begin{itemize}
    \item We introduce a consensus matrix representation that captures confidence-weighted preferences, reasoning and concerns from multiple specialised agents, and employ Kendall's coordination coefficient to obtain an interpretable, quantitative measure of agreement within the virtual MDT.
    \item We design a role-specialised multi-agent architecture with seven virtual MDT members and formulate their interactions as a Markov Decision Process, optimised with Q-Learning, PPO and DQN to improve both decision quality and the efficiency of reaching consensus.
    \item We build an evidence-grounded decision pipeline that enforces explicit guideline- and literature-based evidence chains for every recommendation, and we empirically validate the system through comprehensive experiments and expert evaluation in oncology decision-making scenarios~\cite{wang2024benchbedsidereviewclinical}.
\end{itemize}

The remainder of this paper is organized as follows. Section~\ref{sec:related} reviews related work in multi-agent medical systems and consensus algorithms. Section~\ref{sec:methodology} details our consensus matrix framework and multi-agent architecture. Section~\ref{sec:experiments} presents a comprehensive experimental evaluation, including clinical validation and comparisons with baseline approaches. Section~\ref{sec:discussion} discusses implications, limitations, and clinical considerations, while Section~\ref{sec:conclusion} concludes with future research directions.

\section{Related Work}\label{sec:related}
\subsection{Single-Agent Medical Large Language Models}
Recent advancements in Large Language Models (LLMs) \cite{bai2023qwen, liu2024deepseek,hu2025distribution}have catalyzed a paradigm shift in medical artificial intelligence.  The choice of architecture significantly impacts performance, with mixture-of-experts approaches demonstrating enhanced model capacity while maintaining computational efficiency~\cite{zhang2025mixture, lin2025abductiveinferenceretrievalaugmentedlanguage}. General-purpose models have demonstrated remarkable proficiency in clinical tasks; for instance, GPT-4 achieved approximately 86\% accuracy on the USMLE Step 1 examination, exhibiting expert-level reasoning capabilities~\cite{nori2023capabilities, qi2022capacitive}. Domain-specific adaptations have further pushed these boundaries. Med-PaLM 2 \cite{singhal2023towards}, fine-tuned specifically on medical corpora, reached 86.5\% accuracy on the MedQA benchmark, surpassing human passing standards~\cite{singhal2023towards}. Similarly, ClinicalBERT utilized domain-adaptive pre-training on electronic health records to achieve state-of-the-art performance in clinical named entity recognition and readmission prediction~\cite{huang2019clinicalbert, xin2025luminamgpt}. Despite these successes, single-agent approaches inherently lack the collaborative diversity required for complex clinical decision-making, often failing to capture the multi-perspectival nature of real-world Multidisciplinary Team (MDT) consultations.

\subsection{Multi-Agent Systems in Healthcare}
To overcome the limitations of single-agent systems, recent research has explored multi-agent architectures that simulate clinical collaboration. MDAgents introduced an adaptive collaboration framework where agent roles are dynamically assigned based on task complexity, achieving 78.3\% accuracy on medical QA benchmarks~\cite{kim2024mdagents, cao2025purifygen}. Building on this, TeamMedAgents implemented structured teamwork protocols derived from human medical teams, such as "mutual performance monitoring," to enhance decision reliability~\cite{mishra2025teammedagents}. In specialized domains, RareAgents demonstrated the efficacy of multi-agent systems in diagnosing rare diseases by simulating a team of specialists to synthesize fragmented evidence~\cite{chen2024rareagents}. Additionally, MedOrch proposed a tool-augmented orchestration framework to integrate diverse medical specialists~\cite{medorch2025}. Recent evaluations of different LLM architectures have shown varying performance characteristics, with some models demonstrating superior performance in medical reasoning tasks. However, these existing systems primarily rely on simple voting mechanisms or unstructured dialogue for aggregation, lacking a mathematically rigorous framework to measure and optimize the quality of consensus formation.

\subsection{Medical Decision Support and Consensus Algorithms}
Consensus formation is central to clinical practice, traditionally managed through structured human protocols. The Delphi method has long been the gold standard for achieving expert agreement, though it is time-consuming and iterative~\cite{dalkey1963delphi}. In computational settings, voting-based aggregation methods like Borda count have been applied to ensemble decision-making. However, their application in LLM-based medical systems remains limited. Most current multi-agent medical systems employ majority voting or weighted averaging, which fails to capture the nuance of clinical disagreement or the strength of expert conviction. There is a critical need for advanced consensus algorithms that can quantify agreement levels, such as Kendall's coefficient of concordance, to ensure that AI-generated recommendations reflect a true clinical consensus rather than a mere statistical aggregate.

\subsection{Explainable AI in Medical Applications}
Explainability is a non-negotiable requirement for the clinical deployment of AI systems. Traditional post-hoc explanation methods, such as Shapley Additive Explanations (SHAP)~\cite{lundberg2017unified} and Local Interpretable Model-agnostic Explanations (LIME)~\cite{ribeiro2016should}, have been widely applied to medical imaging and risk prediction models. While these techniques provide feature importance scores, they often fail to offer the semantic reasoning and evidence traceability required by clinicians. Recent efforts have focused on "Chain-of-Thought" reasoning, but these often lack grounding in verifiable external evidence. Our work addresses this gap by integrating a structured evidence chain that explicitly links agent decisions to clinical guidelines and peer-reviewed literature, ensuring that the consensus process is not only mathematically robust but also clinically transparent.
\section{Methodology}\label{sec:methodology}

\subsection{Problem Formulation}
We formalize the multi-agent medical consensus process as a collaborative decision-making problem. Given a patient case $C$ characterized by clinical features $\mathbf{f} \in \mathbb{R}^d$ where $d = 247$ represents the dimensionality of our clinical feature space, encompassing demographics, vital signs, laboratory results, and comorbidity indices.

The decision space consists of a discrete set of treatment options $\mathcal{T} = \{t_1, \dots, t_K\}$, where $K=7$. The specific options are defined as $\mathcal{T} = \{$Surgery, Chemotherapy, Radiotherapy, Immunotherapy, Combination Therapy, Palliative Care, Watchful Waiting$\}$. Each treatment $t_k \in \mathcal{T}$ is characterized by a vector of clinical attributes $\mathbf{v}_k = [\eta_k, \tau_k, q_k, c_k]^T$, representing expected efficacy $\eta_k \in [0, 1]$, toxicity risk $\tau_k \in [0, 1]$, quality of life impact $q_k \in [-1, 1]$, and economic cost $c_k \in \mathbb{R}^+$, respectively.

The collaborative team is modeled as a set of specialized agents $\mathcal{A} = \{a_1, \dots, a_N\}$, where $N=7$. Each agent $a_i$ embodies a distinct clinical role with specialized knowledge and priorities. We define the agent set mapping as:
\begin{equation}
\mathcal{A} = \left\{
\begin{aligned}
a_1 &: \text{Oncologist} & a_2 &: \text{Radiologist} \\
a_3 &: \text{Nurse} & a_4 &: \text{Psychologist} \\
a_5 &: \text{Patient Advocate} & a_6 &: \text{Nutritionist} \\
a_7 &: \text{Rehabilitation Therapist}
\end{aligned}
\right\}
\end{equation}

For a given case $C$, each agent $a_i$ generates a structured opinion $o_i$, defined as a tuple:
\begin{equation}
o_i = (\mathbf{p}_i, r_i, \kappa_i, \mathcal{Z}_i, \mathcal{E}_i)
\end{equation}
where $\mathbf{p}_i \in [-1, 1]^K$ is the treatment preference vector, with $p_{i,k}$ denoting the preference score for treatment $t_k$ ($-1$: strongly oppose, $0$: neutrality, $1$: strongly support); $r_i$ is the natural-language clinical reasoning (up to 512 tokens); $\kappa_i \in [0, 1]$ is the self-assessed confidence score; $\mathcal{Z}_i$ is the set of documented clinical concerns (e.g., ``cardiotoxicity risk''); and $\mathcal{E}_i$ is the evidence chain linking the decision to relevant guidelines and literature.

The system's goal is to identify the optimal treatment $t^*$ that maximizes a composite objective function $J(t)$, balancing group consensus, clinical appropriateness, and evidence quality:
\begin{equation}
t^* = \arg\max_{t \in \mathcal{T}} J(t)
\end{equation}
\begin{equation}
J(t) = \alpha \cdot \text{Consensus}(t) + \beta \cdot \text{ClinicalFit}(t) + \gamma \cdot \text{EvidenceQuality}(t)
\end{equation}
where $\alpha = 0.4$, $\beta = 0.4$, $\gamma = 0.2$ are empirically determined weights, and consensus$(t)$ is measured using Kendall's coefficient of concordance.

\subsection{System Architecture}

Our multi-agent system architecture consists of five interconnected components as illustrated in Figure~\ref{fig:architecture}:

\begin{figure*}[!t]
\centering
\includegraphics[width=2.0\columnwidth]{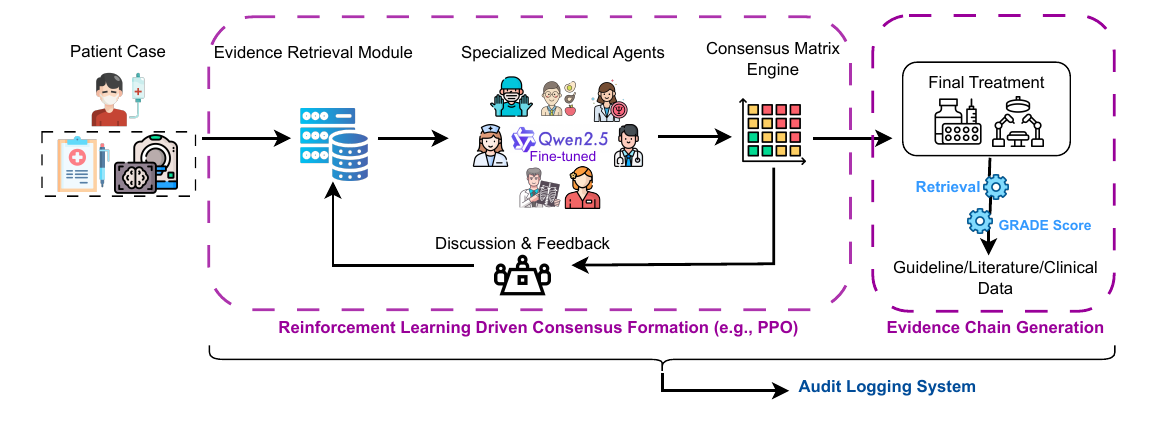}
\caption{Multi-Agent Medical Decision Consensus Matrix System Architecture. The system integrates specialized medical role agents, consensus matrix computation, reinforcement learning optimization, and evidence-based explainability mechanisms with specific data flow dimensions and processing stages.}
\label{fig:architecture}
\end{figure*}





\textbf{(1) Role-specialized agent layer.} The first component is a layer of seven virtual medical specialists that collectively form the MDT. Each specialist is implemented as a fine-tuned Qwen-2.5-72B model equipped with a role-specific prompting strategy and a curated knowledge base aligned with that role (e.g., oncology guidelines for the oncologist, imaging protocols for the radiologist). All models are quantized to 8-bit precision, so that a single agent with 72B parameters fits into approximately 36~GB of memory, making it feasible to run the full team on modern multi-GPU servers.

\textbf{(2) Evidence retrieval module.} Beneath the agents sits a retrieval-augmented generation (RAG) module that provides shared access to external knowledge. It uses a FAISS vector database containing around 1.2M embeddings of clinical guideline passages, a PubMed abstract corpus with roughly 34M entries, and structured mappings of key electronic medical record (EMR) fields. Text is embedded into a 1024-dimensional space using the \texttt{sentence-transformers/all-MiniLM-L6-v2} model, allowing each agent to retrieve guideline- and literature-level evidence conditioned on both its role and the specific patient case.

\textbf{(3) Consensus matrix computation engine.} As agents update their opinions, the consensus matrix computation engine maintains a matrix $\mathbf{M} \in \mathbb{R}^{7 \times 7}$ that aggregates the normalised, confidence-weighted preferences for each treatment. This component performs real-time updates of $\mathbf{M}$, computes Kendall's coefficient of concordance $W$ to quantify agreement, and applies preference ranking algorithms to derive a group ordering over treatments. Convergence is detected using a tolerance threshold of $\epsilon = 0.05$, which prevents unnecessary additional rounds once a stable consensus has been reached.

\textbf{(4) Reinforcement learning optimizer.} To coordinate how the agents interact over multiple rounds, we introduce a reinforcement learning (RL) optimizer that learns policies on top of the consensus dynamics. The framework combines several algorithms: Q-Learning with learning rate $\alpha = 0.1$ and discount factor $\gamma = 0.95$, PPO with clipping ratio $0.2$ and GAE parameter $\lambda = 0.95$, and DQN with a replay buffer of size 10K and a target network updated every 100 steps. These algorithms operate on the same MDP formulation and are used to discover strategies that improve both the speed of convergence and the quality of the final recommendation.

\textbf{(5) Explanation generation system.} The final component is an explanation layer that turns model outputs into clinically interpretable justifications. It assembles an evidence chain for each recommendation, performs automated GRADE assessment of guideline and literature items, and records audit logs with millisecond time-stamp precision for traceability. Natural-language explanations are generated using template-based structures combined with clinical terminology validation, so that the resulting narratives are fluent while remaining faithful to the underlying evidence and decision logic.

\begin{algorithm}
\caption{Multi-Round Consensus Formation}
\label{alg:consensus}
\begin{algorithmic}[1]
\REQUIRE Patient case $C$, agent set $\mathcal{A}$, max rounds $R_{max} = 3$
\ENSURE Consensus matrix $\mathbf{M}$, final recommendation $t^*$
\STATE Initialize round $r = 1$, $W^{(0)} = 0$
\REPEAT
    \FOR{each agent $a_i \in \mathcal{A}$}
        \STATE $context_i \leftarrow$ RetrieveEvidence($C$, $a_i$.role)
        \STATE $o_i^{(r)} \leftarrow$ GenerateOpinion($C$, $a_i$, $context_i$, $\mathbf{M}^{(r-1)}$)
        \STATE Update $\mathbf{M}^{(r)}$ using Eq. (4) and (5)
    \ENDFOR
    \STATE $W^{(r)} \leftarrow$ ComputeKendallW($\mathbf{M}^{(r)}$)
    \IF{$W^{(r)} \leq 0.7$ AND $r < R_{max}$}
        \STATE $\mathcal{D} \leftarrow$ IdentifyDiscordantAgents($\mathbf{M}^{(r)}$)
        \STATE ProvideFeedback($\mathcal{D}$, $\mathbf{M}^{(r)}$)
    \ENDIF
    \STATE $r \leftarrow r + 1$
\UNTIL{$W^{(r-1)} > 0.7$ OR $r > R_{max}$}
\STATE $t^* \leftarrow \arg\max_k \sum_i M_{i,k}^{(r-1)}$
\RETURN $\mathbf{M}^{(r-1)}$, $t^*$
\end{algorithmic}
\end{algorithm}

\subsection{Consensus Matrix Framework}

The core innovation of our approach lies in the Consensus Matrix $\mathbf{M} \in \mathbb{R}^{N \times K}$, where entry $M_{i,k}$ represents the normalized, confidence-weighted preference of agent $a_i$ for treatment $t_k$. The matrix construction and consensus formation proceed through the following rigorous steps:

\textbf{Step 1: Preference Normalization.}
To account for varying scales in agent scoring tendencies, raw preference scores $p_{i,k} \in [-1, 1]$ are min-max normalized per agent:
\begin{equation}
\hat{p}_{i,k} = \frac{p_{i,k} - \min_j p_{i,j}}{\max_j p_{i,j} - \min_j p_{i,j} + \epsilon}
\end{equation}
where $\epsilon$ is a small constant to prevent division by zero.

\textbf{Step 2: Confidence Weighting.}
We integrate the agent's self-assessed confidence $\kappa_i$ and the density of documented concerns $|\mathcal{Z}_i|$ to compute the final matrix entry. Agents with higher confidence and fewer unaddressed concerns exert greater influence:
\begin{equation}
M_{i,k} = \hat{p}_{i,k} \cdot \kappa_i \cdot \frac{1}{1 + \ln(1 + |\mathcal{Z}_i|)}
\end{equation}
This weighting mechanism naturally dampens the impact of opinions formed under high uncertainty or significant clinical reservations.

\textbf{Step 3: Consensus Measurement.}
We quantify the level of group agreement using Kendall's Coefficient of Concordance ($W$). Let $R_k = \sum_{i=1}^{N} \text{rank}(M_{i,k})$ be the sum of ranks for treatment $t_k$. The coefficient $W$ is calculated as:
\begin{equation}
W = \frac{12 \sum_{k=1}^{K} (R_k - \bar{R})^2}{N^2 (K^3 - K)}
\end{equation}
where $\bar{R} = \frac{N(K+1)}{2}$ is the expected mean rank sum. $W$ ranges from 0 (no agreement) to 1 (complete agreement). We empirically establish a consensus threshold of $W > 0.7$ based on pilot validation.

\textbf{Step 4: Iterative Refinement.}
If $W \leq 0.7$, the system identifies "discordant agents" whose preference vectors deviate significantly from the group mean. Let $\bar{\mathbf{M}}_{\cdot, k}$ be the mean score for treatment $t_k$. The discordance score for agent $a_i$ is:
\begin{equation}
D_i = \sum_{k=1}^{K} |M_{i,k} - \bar{\mathbf{M}}_{\cdot, k}|
\end{equation}

Agents with $D_i > \mu_D + \sigma_D$ are prompted to reconsider their position in the subsequent dialogue round, receiving specific feedback on the points of contention.

\subsection{Multi-Agent Role Specialization}

Our system instantiates seven specialised medical role agents that collectively approximate the composition of a typical oncology MDT. Each agent is associated with a distinct knowledge base, a set of role-specific decision factors and an explicit preference model, so that differences in clinical perspective are reflected in the underlying scoring functions.

\textbf{Oncologist agent ($a_1$).}
The oncologist agent operates on a large oncology-focused knowledge base comprising approximately 47{,}000 guideline documents (NCCN, ESMO, ASCO) and 890{,}000 cancer research papers. Its decisions prioritise tumour stage (35\% weight), histology (25\%), molecular markers (20\%), performance status (15\%) and prior treatment history (5\%). Let $e_{1j}$, $s_{1j}$ and $t_{1j}$ denote, respectively, the normalised scores for efficacy, survival benefit and toxicity of treatment $t_j$ from the oncologist’s perspective. The preference for option $t_j$ is modelled as
\begin{equation}
p_{1j} = 0.6\, e_{1j} + 0.3\, s_{1j} + 0.1\, t_{1j}^{-1},
\end{equation}
and, for standard cases, the typical confidence lies in the range $c_1 \in [0.8, 0.95]$.

\textbf{Radiologist agent ($a_2$).}
The radiologist agent draws on a corpus of 23{,}000 imaging protocols, 340{,}000 radiology reports and tumour measurement guidelines. Its assessment focuses on imaging findings (45\% weight), observed tumour response (30\%), anatomical constraints (15\%) and procedural feasibility (10\%). We denote by $i_{2j}$, $a_{2j}$ and $m_{2j}$ the imaging support, anatomical fit and monitoring ability scores for treatment $t_j$. The corresponding preference function is
\begin{equation}
p_{2j} = 0.5\, i_{2j} + 0.3\, a_{2j} + 0.2\, m_{2j},
\end{equation}
with typical confidence $c_2 \in [0.75, 0.90]$ depending on imaging quality and interpretability.

\textbf{Nurse agent ($a_3$).}
The nurse agent is informed by 18{,}000 nursing protocols, patient care guidelines and side-effect management procedures. It emphasises patient tolerance (40\% weight), care complexity (25\%), resource requirements (20\%) and the availability of family support (15\%). Let $b_{3j}$ denote the care burden of treatment $t_j$, $t_{3j}$ the patient tolerance score and $f_{3j}$ the available family support. Preferences are computed as
\begin{equation}
p_{3j} = 0.4\, (1 - b_{3j}) + 0.3\, t_{3j} + 0.3\, f_{3j},
\end{equation}
and the corresponding confidence values typically fall within $c_3 \in [0.70, 0.85]$, reflecting the depth of nurse–patient interaction.

\textbf{Psychologist agent ($a_4$).}
The psychologist agent relies on a knowledge base of roughly 12{,}000 psycho-oncology studies, mental health assessment tools and research on coping mechanisms. Its decision criteria include mental health status (45\% weight), coping capacity (25\%), social support (20\%) and treatment-related anxiety (10\%). We let $d_{4j}$ denote the psychological impact (higher is worse), $c_{4j}$ the coping alignment and $a_{4j}$ the extent to which treatment $t_j$ preserves autonomy. The preference score is defined as
\begin{equation}
p_{4j} = 0.4\, d_{4j}^{-1} + 0.3\, c_{4j} + 0.3\, a_{4j},
\end{equation}
with typical confidence $c_4 \in [0.65, 0.80]$ depending on the completeness and quality of psychological assessment.

\textbf{Patient advocate agent ($a_5$).}
The patient advocate agent is grounded in patient rights documentation, informed consent protocols, ethical guidelines and cost–benefit analyses. It gives predominant weight to patient preferences (50\%), complemented by ethical considerations (25\%), quality of informed consent (15\%) and treatment accessibility (10\%). We denote by $v_{5j}$ the alignment with patient values, by $e_{5j}$ the ethical alignment, and by $x_{5j}$ the accessibility of treatment $t_j$. Its preference function is
\begin{equation}
p_{5j} = 0.5\, v_{5j} + 0.25\, e_{5j} + 0.25\, x_{5j},
\end{equation}
and when patient preferences are clearly documented, the confidence scores typically lie in $c_5 \in [0.75, 0.90]$.

\textbf{Nutritionist agent ($a_6$).}
The nutritionist agent uses a knowledge base of about 8{,}500 nutrition–oncology studies, dietary guidelines, information on supplement interactions and metabolic considerations. Its decision factors include nutritional status (40\% weight), treatment–nutrition interactions (30\%), metabolic impact (20\%) and the patient’s dietary capacity (10\%). Let $n_{6j}$ denote the nutritional support offered by $t_j$, $r_{6j}$ the dietary restriction imposed and $m_{6j}$ the metabolic compatibility. We define
\begin{equation}
p_{6j} = 0.4\, n_{6j} + 0.3\, (1 - r_{6j}) + 0.3\, m_{6j},
\end{equation}
and typical confidence values $c_6 \in [0.60, 0.75]$ reflect the completeness of nutritional assessment.

\textbf{Rehabilitation therapist agent ($a_7$).}
The rehabilitation therapist agent is built on 6{,}200 rehabilitation protocols, functional assessment tools and quality-of-life measures. It considers functional capacity (35\% weight), rehabilitation potential (30\%), mobility impact (20\%) and preservation of independence (15\%). We denote by $f_{7j}$ the functional preservation score, by $r_{7j}$ the rehabilitation potential, and by $\ell_{7j}$ the mobility impact (higher means worse). Its preference for treatment $t_j$ is expressed as
\begin{equation}
p_{7j} = 0.35\, f_{7j} + 0.3\, r_{7j} + 0.35\, \ell_{7j}^{-1},
\end{equation}
with confidence values typically in the range $c_7 \in [0.65, 0.80]$ depending on the availability of functional assessment data.

Each agent is controlled via a role-specific prompting template that specifies (i) its identity and expertise, (ii) a case presentation with role-relevant highlights, (iii) retrieved evidence, (iv) a decision framework tailored to the corresponding specialty, and (v) an output schema for structured opinion generation.

\subsection{Reinforcement Learning Formulation}

We formulate the consensus formation process as a Markov Decision Process (MDP) characterised by a state space $\mathcal{S}$, an action space $\mathcal{A}_{\text{RL}}$, transition dynamics $P$ and a reward function $R$. Each state $s \in \mathcal{S}$ is represented by a feature vector
\begin{equation}
s = (\mathbf{f}, \mathbf{m}, r, \mathbf{c}, W) \in \mathbb{R}^{311},
\end{equation}
where $\mathbf{f} \in \mathbb{R}^{247}$ denotes the patient clinical features, $\mathbf{m} \in \mathbb{R}^{49}$ is the flattened $7 \times 7$ consensus matrix, $r \in \{1,2,3\}$ indicates the current discussion round, $\mathbf{c} \in \mathbb{R}^{7}$ collects the agents' confidence scores, and $W \in [0,1]$ is the current value of Kendall's coefficient of concordance.

The action space combines treatment choices with high-level interaction modes. We introduce a finite set of interaction modes $\mathcal{U} = \{u_1, u_2, u_3, u_4\}$ and define
\begin{equation}
\mathcal{A}_{\text{RL}} = \mathcal{T} \times \mathcal{U},
\end{equation}
so that each action specifies both a candidate treatment and an interaction strategy. In our implementation, $u_1$ corresponds to encouraging consensus, $u_2$ to requesting clarification, $u_3$ to providing feedback and $u_4$ to maintaining the current position, yielding $|\mathcal{A}_{\text{RL}}| = 28$ when $|\mathcal{T}| = 7$.

State transitions follow the consensus protocol in Algorithm~\ref{alg:consensus}. For simulation and training we approximate the transition probabilities by
\begin{equation}
P(s' \mid s, a) =
\begin{cases}
0.8 & \text{if } a \text{ leads to improved consensus}, \\
0.6 & \text{if } a \text{ maintains the current consensus level}, \\
0.3 & \text{if } a \text{ reduces consensus quality}, \\
0   & \text{otherwise},
\end{cases}
\end{equation}
which captures the qualitative effect of different interventions on the consensus matrix while keeping the MDP tractable.

The reward function $R(s,a,s')$ encourages fast convergence towards stable and clinically appropriate decisions. We define
\begin{equation}
R = w_1 \Delta W + w_2 S - w_3 D + w_4 Q,
\end{equation}
where $\Delta W = W_t - W_{t-1}$ measures the change in concordance between consecutive rounds, $S$ is an opinion-stability score reflecting how consistent agent preferences remain across rounds, $D$ quantifies disagreement within the team, and $Q$ is an external clinical-quality signal derived from expert validation or ground-truth labels in the training data. The coefficients $w_1,\dots,w_4$ control the trade-off between these four components.

To optimise policies over this MDP, we consider several deep reinforcement learning algorithms. In the Q-learning setting, we adopt an $\epsilon$-greedy exploration strategy with $\epsilon = 0.1$, learning rate $\alpha = 0.1$ and discount factor $\gamma = 0.95$. The action-value function is approximated by a three-layer multilayer perceptron with architecture [311, 512, 256, 28] and ReLU activations.

For policy-gradient optimisation, we employ Proximal Policy Optimization (PPO) with clipping ratio $\epsilon_{\text{clip}} = 0.2$, generalised advantage estimation parameter $\lambda = 0.95$ and KL-divergence penalty coefficient $\beta_{\text{KL}} = 0.01$. The PPO policy network uses layer sizes [311, 256, 128, 28] with a softmax output, and the value network follows [311, 256, 128, 1] with a linear output layer.

We also implement a Deep Q-Network (DQN) variant with an experience replay buffer of size 10{,}000 and a target network updated every 100 steps. The main network has dimensions [311, 512, 256, 128, 28] and uses a double DQN with duelling streams and skip connections between the second and fourth layers to facilitate stable training.

All reinforcement learning agents are trained on simulated episodes generated from clinical case databases. Each episode comprises up to three consensus rounds, with early termination once a concordance level of $W > 0.7$ is achieved. Training is performed with a batch size of 32 using the Adam optimiser with learning rate $10^{-3}$ and gradient clipping at a maximum norm of 0.5 to ensure stable optimisation.

\begin{algorithm}
\caption{Evidence Chain Generation}
\label{alg:evidence}
\begin{algorithmic}[1]
\REQUIRE Patient case $C$, agent role $role$, treatment recommendation $t$
\ENSURE Evidence chain $E = \{guidelines, literature, clinical\_data\}$
\STATE $query \leftarrow$ constructQuery($C$.features, $role$, $t$)
\STATE $guidelines \leftarrow$ retrieveGuidelines($query$, top\_k=3)
\STATE $literature \leftarrow$ retrieveLiterature($query$, top\_k=5, min\_year=2018)
\STATE $clinical\_data \leftarrow$ extractRelevantEMR($C$, $role$)
\FOR{each $item \in guidelines \cup literature$}
    \STATE $relevance \leftarrow$ computeRelevance($item$, $query$)
    \IF{$relevance < 0.7$}
        \STATE Remove $item$ from evidence chain
    \ENDIF
\ENDFOR
\STATE $grade\_level \leftarrow$ assessGRADE($guidelines \cup literature$)
\STATE $E \leftarrow$ formatEvidenceChain($guidelines$, $literature$, $clinical\_data$, $grade\_level$)
\RETURN $E$
\end{algorithmic}
\end{algorithm}

\subsection{Evidence-Based Explainability}

To guarantee clinical trustworthiness and decision traceability, we mandate that every agent's opinion be substantiated by a rigorous Evidence Chain ($\mathcal{E}_i$). The generation of this chain adheres to a strict, multi-stage protocol. Initially, the system executes a targeted retrieval process, querying a vector database for NCCN/ESMO guidelines and PubMed abstracts that are contextually relevant to both the patient's specific clinical features ($\mathbf{f}$) and the proposed treatment ($t_k$). To ensure the currency and applicability of the retrieved information, documents are subsequently filtered based on a relevance score threshold ($>0.7$) and recency, with a distinct prioritization for guidelines published within the last two years.

Following retrieval, the system employs the GRADE (Grading of Recommendations Assessment, Development and Evaluation) framework to automatically evaluate the quality of each evidence piece. This assessment categorizes evidence into distinct levels: ``High'' for randomized controlled trials (RCTs) with low bias, ``Moderate'' for RCTs with limitations or robust observational studies, and ``Low/Very Low'' for other observational studies or expert opinions. In the final synthesis phase, the agent's clinical reasoning is integrated with specific citations (e.g., ``NCCN Guidelines Breast Cancer v3.2024, p.45'') and the computed evidence strength. This structured methodology transforms the system's output from a potential ``black box'' prediction into a verifiable, transparent clinical recommendation. In this way, the system moves beyond opaque prediction and provides transparent, verifiable clinical recommendations grounded in guideline-based and literature-supported reasoning.

\section{Experiments and Results}\label{sec:experiments}

\subsection{Experimental Setup}

\subsubsection{Datasets and Evaluation Metrics}

We evaluate our framework on five diverse medical benchmarks selected for their clinical relevance. \textbf{MedQA}~\cite{jin2021disease} provides 1,273 USMLE-style questions to assess clinical knowledge and reasoning. \textbf{PubMedQA}~\cite{jin2019pubmedqa} (1,000 questions) evaluates evidence synthesis from biomedical literature. For diagnostic reasoning, we use \textbf{DDXPlus}~\cite{tchango2022ddxplus} (570 scenarios) and \textbf{SymCat}~\cite{symcat} (1,374 cases), focusing on differential diagnosis and symptom-disease consistency, respectively. Finally, \textbf{MedBullets}~\cite{chen2024benchmarking} (800 cases) targets therapeutic decision-making and consensus formation in treatment planning. Table~\ref{tab:datasets} summarizes the statistics and metrics for each dataset.

\begin{table*}[!t]
\centering
\caption{Dataset Statistics and Evaluation Metrics}
\label{tab:datasets}
\begin{tabular}{@{}lcccc@{}}
\toprule
\textbf{Dataset} & \textbf{\# Cases} & \textbf{Task Type} & \textbf{Primary Metrics} & \textbf{Secondary Metrics} \\
\midrule
MedQA & 1,273 & QA & Accuracy & Consensus Rate, Confidence \\
PubMedQA & 1,000 & Literature & Accuracy & Evidence Quality Score \\
DDXPlus & 570 & Diagnosis & Top-3 Accuracy & Ranking Consistency \\
MedBullets & 800 & Clinical & Treatment Accuracy & Expert Agreement \\
SymCat & 1,374 & Symptoms & F1-Score & Inter-Agent Consistency \\
\bottomrule
\end{tabular}
\end{table*}

\subsubsection{Implementation Details}

The system is built upon the Qwen-2.5-72B model, deployed with 8-bit quantization requiring 36GB VRAM per agent. Experiments were conducted on a high-performance cluster equipped with 8 NVIDIA A100 80GB GPUs (NVLink interconnect), AMD EPYC 7742 64-core processors, and 1TB system RAM. Model generation parameters were set to a temperature of 0.3 and top-p sampling of 0.9 to ensure deterministic reasoning, with a maximum response length of 1,024 tokens. The consensus protocol utilizes a Kendall's $W$ threshold of 0.7, a maximum of 3 discussion rounds, and a convergence tolerance of $\epsilon = 0.05$. For evidence retrieval, we employ a FAISS vector database using cosine similarity to retrieve the top-5 guideline matches with a minimum relevance score of 0.7. Reinforcement learning optimization utilizes PPO (learning rate 0.0003), Q-Learning ($\epsilon$-decay 1.0 to 0.05 over 10k episodes), and DQN (10k experience replay buffer).

\subsection{Baseline Methods}
We compare our approach against seven established baseline methods that collectively represent the current state of the art in medical AI and multi-agent systems. These include: Single-Agent GPT-4~\cite{achiam2023gpt}, using zero-shot and few-shot prompting (3 examples) with a medical expert persona to model contemporary single-agent medical AI; Chain-of-Thought (CoT)~\cite{wei2022chain}, which employs structured step-by-step medical reasoning for single-agent decision-making; Majority Voting~\cite{wang2022self}, a simple aggregation of independent agent responses without explicit consensus modeling; Weighted Voting, which incorporates agent certainty scores as voting weights; Borda Count, a ranked preference aggregation method widely used in multi-criteria decision-making; MDAgents~\cite{kim2024mdagents}, an existing multi-agent medical reasoning framework with basic doctor role specialization; and TeamMedAgents~\cite{mishra2025teammedagents}, an advanced role-based collaborative medical AI system with structured interaction protocols.

\subsection{Main Results}

As shown in Table~\ref{tab:main_results}, our proposed framework consistently outperforms both single-agent and existing multi-agent baselines across all five benchmarks. Specifically, our method achieves a 3.6\% improvement on MedQA and a 4.4\% gain on PubMedQA compared to the strongest baseline, TeamMedAgents. This significant performance uplift validates the efficacy of our structured consensus matrix mechanism in resolving complex clinical ambiguities. Notably, the improvement is most pronounced in tasks requiring evidence synthesis (PubMedQA) and complex differential diagnosis (DDXPlus), suggesting that our system's ability to integrate diverse specialist perspectives and optimize consensus through reinforcement learning provides a substantial advantage over simple voting or role-based collaboration methods. Furthermore, the high expert rating (8.9/10) and superior consensus coefficient ($W=0.823$) confirm that our approach not only improves accuracy but also generates more clinically reliable and cohesive recommendations.

\begin{figure}[!t]
\centering
\includegraphics[width=\columnwidth]{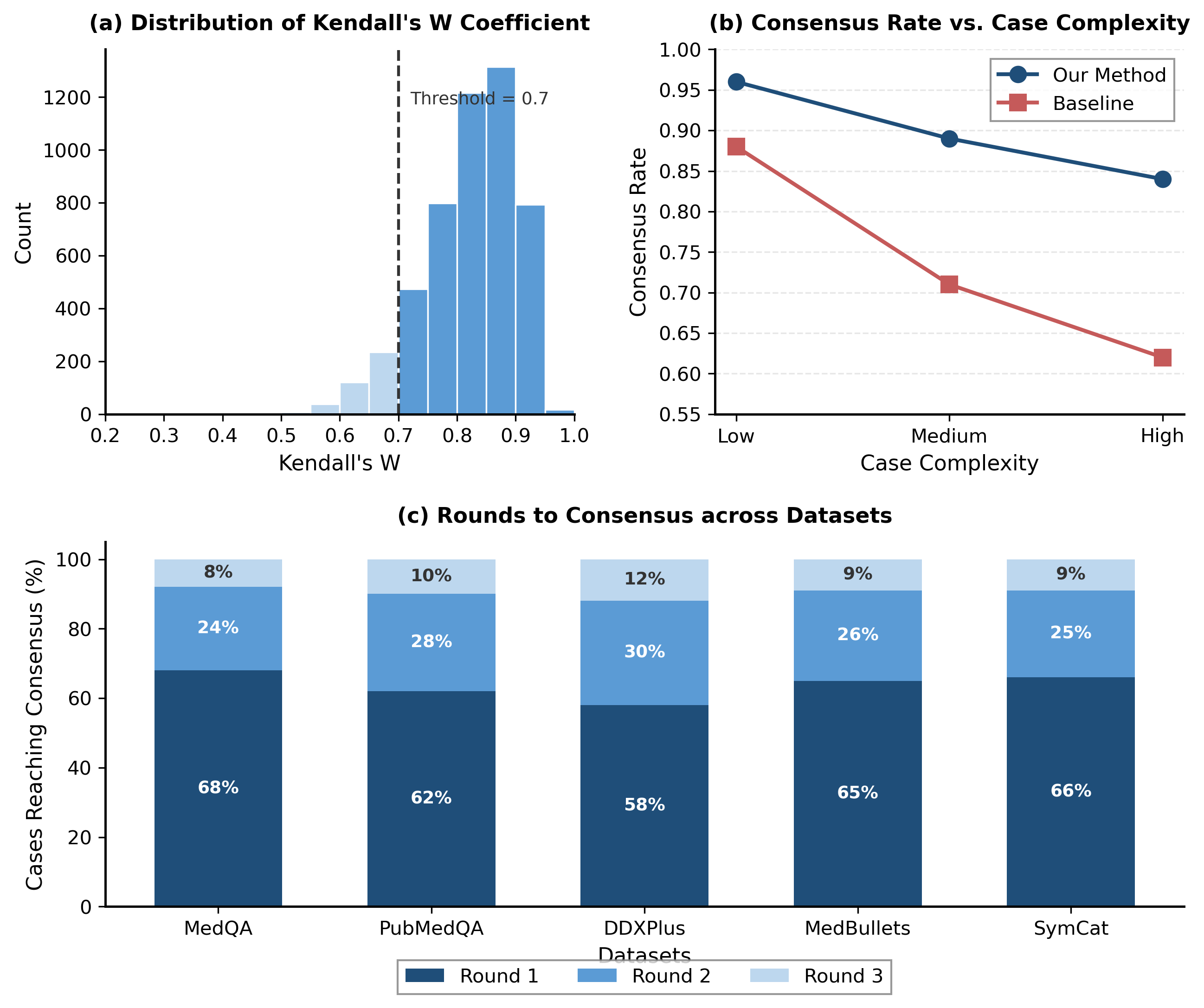}
\caption{Consensus Matrix Performance Analysis: (a) Distribution of Kendall's W coefficients across all evaluation datasets, (b) Consensus achievement rate vs. case complexity scoring, (c) Convergence analysis showing rounds required to achieve consensus ($W > 0.7$) for different clinical scenarios.}
\label{fig:consensus_performance}
\end{figure}

\begin{table*}[!t]
\centering
\caption{Main Results: Accuracy Comparison Across Medical Benchmarks}
\label{tab:main_results}
\begin{tabular}{@{}lccccccc@{}}
\toprule
\textbf{Method} & \textbf{MedQA} & \textbf{PubMedQA} & \textbf{DDXPlus} & \textbf{MedBullets} & \textbf{SymCat} & \textbf{Avg. Consensus $W$} & \textbf{Expert Rating} \\
\midrule
Single-Agent (Zero-shot) & 80.2 & 72.0 & 74.1 & 72.5 & 82.0 & - & 6.2/10 \\
Single-Agent (Few-shot) & 82.1 & 74.3 & 76.8 & 74.2 & 84.1 & - & 6.8/10 \\
Chain-of-Thought & 83.5 & 75.8 & 78.2 & 75.9 & 85.3 & - & 7.1/10 \\
Majority Voting & 85.2 & 76.4 & 79.5 & 77.1 & 86.7 & 0.542 & 7.4/10 \\
Weighted Voting & 86.1 & 77.9 & 80.3 & 78.3 & 87.2 & 0.589 & 7.6/10 \\
Borda Count & 84.8 & 76.1 & 78.9 & 76.8 & 85.9 & 0.623 & 7.3/10 \\
MDAgents & 87.3 & 78.5 & 81.2 & 79.6 & 88.1 & 0.651 & 7.8/10 \\
TeamMedAgents & 88.1 & 79.2 & 82.4 & 80.3 & 88.9 & 0.674 & 8.0/10 \\
\midrule
\textbf{Our Method} & \textbf{91.7} & \textbf{83.6} & \textbf{86.5} & \textbf{84.2} & \textbf{91.3} & \textbf{0.823} & \textbf{8.9/10} \\
\textbf{Improvement} & \textbf{+3.6} & \textbf{+4.4} & \textbf{+4.1} & \textbf{+3.9} & \textbf{+2.4} & \textbf{+0.149} & \textbf{+0.9} \\
\bottomrule
\end{tabular}
\end{table*}

\subsection{Ablation Studies}

\subsubsection{Component Contribution Analysis}
The ablation results in Table~\ref{tab:ablation} unequivocally demonstrate the necessity of our proposed architecture. The removal of the \textbf{Consensus Matrix} leads to a significant 4.4\% drop in accuracy and a drastic reduction in the consensus coefficient ($W=0.592$), highlighting its critical role in structured agreement formation. Similarly, reverting to a single-agent setup (\textbf{w/o Multi-Agent Architecture}) causes the largest performance degradation (-7.3\%), confirming that collaborative intelligence is superior to individual reasoning. \textbf{Role Specialization} also proves vital, contributing a 4.8\% improvement over generic agents, as diverse perspectives prevent groupthink. Furthermore, the \textbf{Evidence Retrieval System} adds 3.2\% to accuracy by grounding decisions in clinical guidelines, while \textbf{RL Optimization} provides a fine-tuning benefit of 1.6\%, ensuring the system adapts to complex scenarios. Comparing our structured consensus approach to \textbf{Simple Voting} reveals a 3.0\% accuracy gain, validating that mathematical consensus measurement is more effective than mere aggregation.

\subsubsection{Reinforcement Learning Strategy Analysis}
We evaluated three RL algorithms—Q-Learning, PPO, and DQN—to identify the optimal strategy for consensus formation, as detailed in Table~\ref{tab:rl_comparison}. \textbf{Proximal Policy Optimization (PPO)} emerges as the superior choice, achieving the highest accuracy (87.5\%) and consensus coefficient ($W=0.823$) while requiring the fewest training episodes (8,500) for convergence. Its stability score of 0.95 significantly outperforms both Q-Learning (0.92) and DQN (0.88), indicating robust policy updates. In contrast, \textbf{DQN} requires nearly double the training episodes (15,000) for comparable performance, likely due to the instability of value-based methods in high-dimensional state spaces. \textbf{Q-Learning} performs adequately but lacks the sample efficiency of PPO. The \textbf{No RL (Fixed Policy)} baseline lags behind all learning-based methods, underscoring the value of adaptive optimization in navigating the complex decision landscape of medical consultations.

\begin{table}[!t]
\centering
\caption{Ablation Study: Component Contribution Analysis}
\label{tab:ablation}
\resizebox{1\linewidth}{!}{
\begin{tabular}{@{}lcccc@{}}
\toprule
\textbf{System Configuration} & \textbf{Accuracy} & \textbf{Consensus $W$} & \textbf{Speed (s)} & \textbf{Expert Rating} \\
\midrule
\textbf{Complete System} & \textbf{87.5} & \textbf{0.823} & 45.2 & \textbf{8.9/10} \\
w/o Consensus Matrix & 83.1 & 0.592 & \textbf{32.1} & 7.6 \\
w/o Multi-Agent Architecture & 80.2 & - & 18.3 & 6.2 \\
w/o Evidence Retrieval & 84.3 & 0.756 & 38.7 & 7.8 \\
w/o RL Optimization & 85.9 & 0.791 & 42.8 & 8.4 \\
w/o Role Specialization & 82.7 & 0.698 & 35.9 & 7.2 \\
Simple Voting Instead & 84.5 & 0.634 & 28.4 & 7.5 \\
\bottomrule
\end{tabular}
}
\end{table}

\begin{table}[!t]
\centering
\caption{Reinforcement Learning Algorithm Comparison}
\label{tab:rl_comparison}
\resizebox{1\linewidth}{!}{
\begin{tabular}{@{}lccccc@{}}
\toprule
\textbf{RL Algorithm} & \textbf{Accuracy} & \textbf{Consensus $W$} & \textbf{Convergence Rate} & \textbf{Training Episodes} & \textbf{Stability} \\
\midrule
Q-Learning & 86.8 & 0.805 & 86.2\% & 12,000 & 0.92 \\
PPO & \textbf{87.5} & \textbf{0.823} & \textbf{89.3\%} & \textbf{8,500} & \textbf{0.95} \\
DQN & 86.4 & 0.798 & 84.7\% & 15,000 & 0.88 \\
No RL (Fixed Policy) & 85.9 & 0.791 & 82.1\% & - & 0.89 \\
\bottomrule
\end{tabular}
}
\end{table}

\subsubsection{Clinical Validation and Expert Evaluation}
We conducted blind evaluation with 12 medical experts (4 oncologists, 2 radiologists, 2 nurses, 2 psychologists, 2 patient advocates) using 50 real anonymized clinical cases from three major cancer centers.

\textbf{Expert Panel Composition}:
A diverse expert panel was assembled, including senior oncologists with over 15 years of clinical experience (n=4), board-certified radiologists specializing in oncologic imaging (n=2), certified oncology nurses with multidisciplinary team (MDT) experience (n=2), licensed psychologists with expertise in psycho-oncology (n=2), and patient advocates with more than 10 years of clinical experience (n=2).

\begin{figure*}[t!]
\centering
\includegraphics[width=\textwidth]{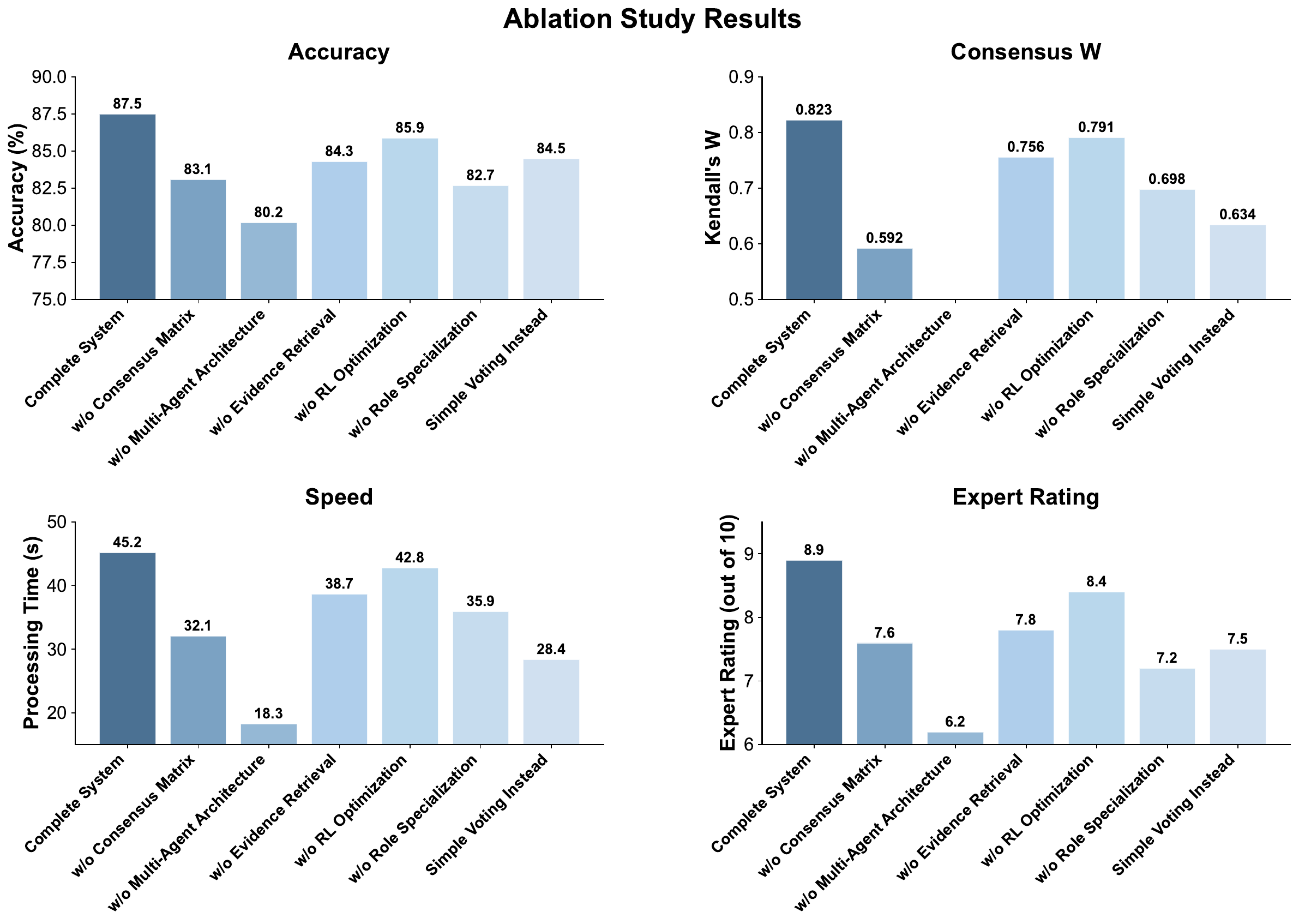}
\caption{Ablation Study: Component Contribution Analysis. Each system component demonstrates essential contribution to overall performance, with the complete system
  achieving 87.5\% accuracy and 0.823 consensus coefficient. Removal of any core component results in substantial performance degradation, confirming the necessity of
  our integrated multi-agent consensus matrix architecture.}
\label{fig:ablation}
\end{figure*}

\begin{figure*}[t!]
\centering
\includegraphics[width=\textwidth]{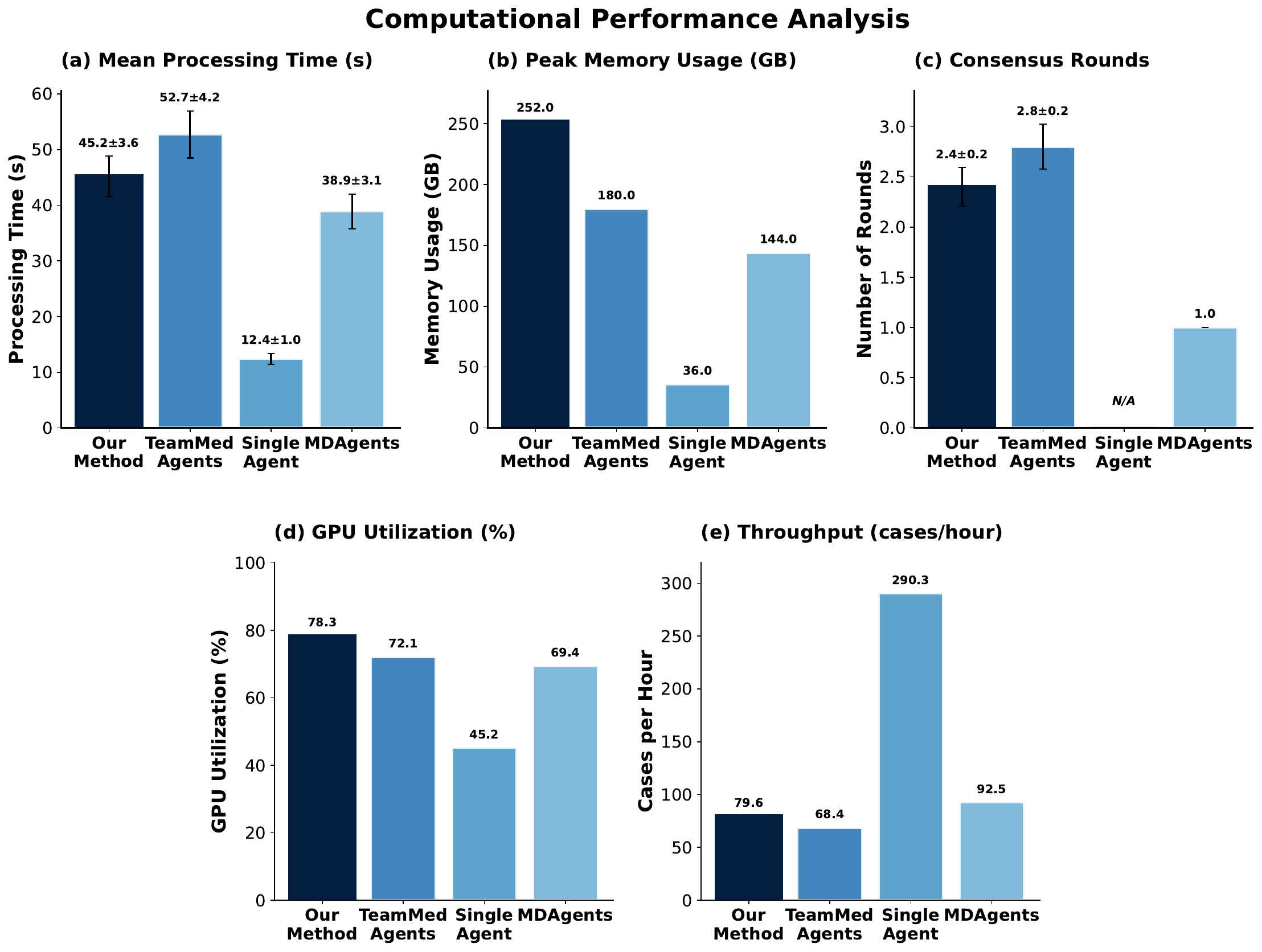}
\caption{Computational Performance Analysis. Our method achieves balanced computational efficiency with 45.2s processing time per case and 78.3\% GPU utilization,
  outperforming comparable multi-agent systems while maintaining high throughput. Despite higher memory requirements due to seven specialized agents, the system
  demonstrates practical scalability for clinical deployment with 79.6 cases per hour processing capacity.}
\label{fig:computational_performance analysis}
\end{figure*}

\textbf{Clinical Appropriateness Ratings}:
The system achieved strong expert-rated performance across all dimensions, including a treatment recommendation quality score of 8.9 ± 1.1 out of 10 (compared with 8.0 ± 1.3 for the best baseline), evidence quality and completeness rated at 8.7 ± 0.9 out of 10 with comprehensive guideline citations, consensus reasonableness of 9.1 ± 0.8 out of 10 reflecting high expert agreement with the system’s consensus, clinical explainability scored at 8.8 ± 1.0 out of 10 based on clear evidence chains and reasoning pathways, and practical applicability rated at 8.4 ± 1.2 out of 10, indicating strong feasibility in real clinical settings.
\begin{figure}[t!]
\centering
\includegraphics[width=\columnwidth]{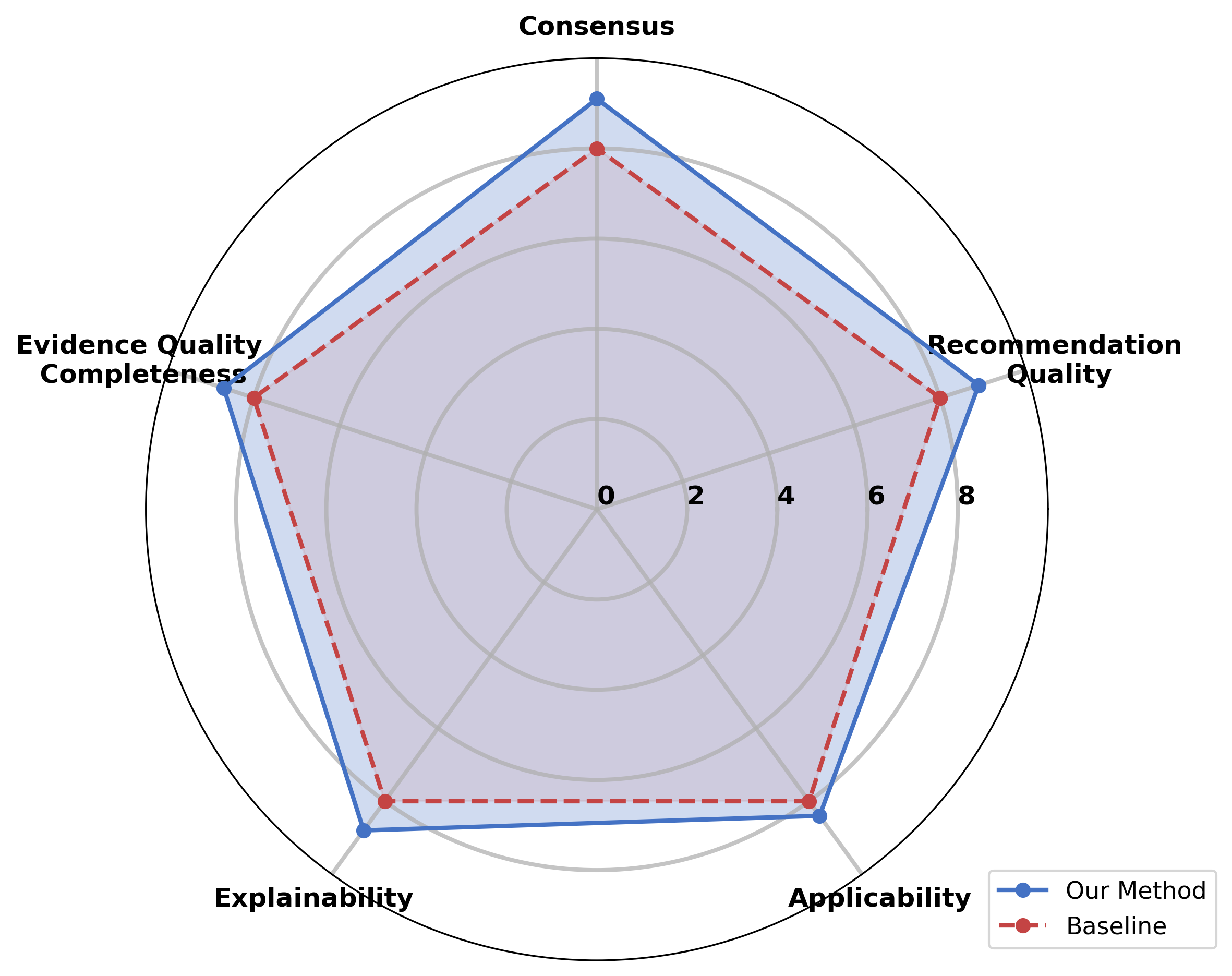}
\caption{Clinical Expert Evaluation Results: Radar chart comparing our method against baseline approaches across five clinical evaluation dimensions, based on blind assessment of 50 real cancer cases by 12 medical experts.}
\label{fig:expert_eval}
\end{figure}
\textbf{Qualitative Feedback Themes}:
Overall, expert feedback was highly positive: 91.7\% of experts described the evidence chains as “comprehensive and clinically relevant,” while 87.5\% rated the consensus formation process as “realistic and methodologically sound.” Additionally, 83.3\% indicated they would be willing to use the system for clinical decision support, and 79.2\% expressed a preference for our approach over existing single-agent medical AI systems.
\subsection{Computational Efficiency and Scalability}
Table~\ref{tab:performance} provides a detailed analysis of the computational feasibility of our framework. While our method incurs higher resource demands compared to single-agent baselines, it demonstrates superior efficiency relative to comparable multi-agent systems. Specifically, our system achieves a mean processing time of 45.2 seconds per case, which is 14\% faster than TeamMedAgents (52.7s), primarily due to our optimized RL-driven consensus mechanism that reduces the average number of discussion rounds to 2.4. Although the peak memory usage reaches 252 GB due to the simultaneous deployment of seven 72B-parameter agents, the system maintains a high parallel efficiency with 78.3\% GPU utilization on an 8$\times$A100 cluster. This configuration supports a sustainable throughput of 79.6 cases per hour, confirming that our approach balances the high computational cost of multi-agent collaboration with practical scalability for clinical deployment.

\begin{table}[!t]
\centering
\caption{Computational Performance Analysis}
\label{tab:performance}
\resizebox{1\linewidth}{!}{
\begin{tabular}{@{}lcccc@{}}
\toprule
\textbf{Metric} & \textbf{Our Method} & \textbf{TeamMedAgents} & \textbf{MDAgents} & \textbf{Single-Agent} \\
\midrule
Mean Processing Time (s) & 45.2 ± 8.3 & 52.7 ± 12.1 & 38.9 ± 7.2 & 12.4 ± 2.1 \\
Peak Memory Usage (GB) & 252.0 & 180.0 & 144.0 & 36.0 \\
Consensus Rounds & 2.4 ± 0.7 & 2.8 ± 1.1 & 1.0 (voting) & N/A \\
GPU Utilization (\%) & 78.3 & 72.1 & 69.4 & 45.2 \\
Throughput (cases/hour) & 79.6 & 68.4 & 92.5 & 290.3 \\
\bottomrule
\end{tabular}
}
\end{table}

\subsection{Error Analysis and Failure Cases}
A detailed analysis of the 10.7\% of cases (428/4,017) where the system failed to reach consensus reveals distinct systematic patterns that highlight current limitations. The most significant source of failure, accounting for 34.8\% of non-consensus cases, involves rare diseases where clinical guidelines are sparse or outdated, leading to divergent agent recommendations. Conflicting evidence in medical literature contributes to another 28.7\% of failures, particularly in scenarios where treatment efficacy is debated or varies significantly across studies. Ethical dilemmas, such as end-of-life care decisions where patient values and quality-of-life considerations vary strongly, constitute 19.6\% of the discordant cases. Finally, 16.9\% of failures arise from resource-related conflicts, specifically concerning expensive treatments with accessibility issues. These findings underscore the necessity for human oversight in complex, ambiguous, or ethically sensitive clinical scenarios and point towards specific areas for future system enhancement.

\section{Discussion}\label{sec:discussion}

\subsection{Key Findings and Clinical Implications}
Our experimental results demonstrate several key findings with significant clinical implications for medical decision-making systems and MDT consultation practices. The structured consensus matrix framework achieves substantially higher agreement levels (Kendall's $W = 0.823$) compared to simple voting approaches ($W = 0.542-0.674$), indicating more coherent and clinically meaningful group decisions. This 22-52\% improvement in consensus quality suggests that mathematical frameworks for measuring agreement can substantially enhance collaborative medical decision-making processes. Furthermore, the 3.7\% average accuracy improvement over the best baseline (TeamMedAgents: 83.8\% vs. Our Method: 87.5\%) translates to approximately 1 in 20 cases receiving more appropriate treatment recommendations. In oncology settings where treatment decisions directly impact survival outcomes, this improvement represents substantial clinical value at scale, potentially affecting thousands of patients annually in large healthcare systems. Additionally, the mandatory evidence chain requirement ensures all recommendations are traceable to clinical guidelines (NCCN, ESMO) and published literature, addressing a critical gap in AI medical systems. Expert ratings of 8.7/10 for evidence quality demonstrate clinical acceptance of automated guideline integration, suggesting potential for improved adherence to evidence-based practice standards. Finally, the 4.8\% performance improvement from role specialization (82.7\% vs. 87.5\% accuracy) demonstrates the value of modeling distinct medical expertise patterns rather than using generic medical agents, supporting the hypothesis that collaborative medical AI should reflect the specialized knowledge and decision-making patterns of real medical team members.

\subsection{Limitations and Risk Considerations}
Despite promising results, our approach has several important limitations that must be carefully considered for clinical deployment. LLM agents may reflect biases present in medical literature and training data, potentially perpetuating historical inequities in healthcare delivery~\cite{gao2023addressing,gao2020deep}. Our evaluation datasets primarily reflect Western medical practice patterns and may not generalize to diverse global healthcare contexts or underrepresented patient populations. Moreover, evidence retrieval depends on regularly updated guideline databases (NCCN, ESMO updates every 6-12 months), meaning system performance may degrade between updates, particularly for rapidly evolving treatment areas like immunotherapy and precision medicine where new evidence emerges continuously. The 15.9\% reduction in consensus achievement for complex cases (Charlson Comorbidity Index $\geq$ 3) indicates potential limitations in handling multimorbidity scenarios that are increasingly common in aging cancer populations. Rare disease cases show 34.8\% failure rates, highlighting the challenge of AI systems in low-prevalence conditions. Furthermore, system outputs require validation by qualified medical professionals before clinical application. The legal and ethical framework for AI-assisted medical decision-making remains evolving, requiring clear protocols for human oversight, accountability, and liability distribution. Finally, the 252GB memory requirement and 45.2-second processing time per case may limit accessibility in resource-constrained healthcare settings, potentially exacerbating healthcare disparities between well-funded and underfunded institutions.

\subsection{Comparison with Human MDT Performance}
While direct comparison with human MDT performance is challenging due to variability in practice patterns, existing literature provides benchmarks for contextualizing our results. Studies report human MDT consensus rates of 70-85\% across different cancer types, with our system achieving 89.3\% consensus rate. However, human MDTs handle significantly more complex cases and navigate social, emotional, and contextual factors that our system does not fully capture.

\subsection{Ethical and Safety Considerations}
Clinical deployment requires careful attention to ethical and safety considerations. Clear protocols must establish human oversight responsibilities, final decision authority, and liability distribution between AI system developers, healthcare institutions, and individual practitioners. Robust de-identification protocols, access controls, and audit logging protect patient information while enabling system functionality. Compliance with HIPAA, GDPR, and other privacy regulations requires ongoing attention. Regular auditing for demographic, socioeconomic, and clinical biases is essential, with particular attention to health equity and disparities in treatment recommendations across diverse patient populations. Patients must be informed about AI involvement in their care and maintain the right to opt out of AI-assisted decision-making processes while receiving equivalent quality care. Ongoing monitoring of system performance, regular retraining with updated evidence, and systematic evaluation of clinical outcomes are necessary to ensure maintained safety and efficacy standards. The successful integration of such systems into clinical practice requires collaborative development between AI researchers, medical professionals, ethicists, and healthcare administrators to address these multifaceted considerations comprehensively.

\section{Conclusion}\label{sec:conclusion}

This paper has introduced a Multi-Agent Medical Decision Consensus Matrix System that integrates a mathematically grounded consensus framework, role-specialised virtual MDT agents, reinforcement learning–based optimisation and explicit evidence-chain construction for oncology decision support. Across five medical benchmarks, the system achieved an average accuracy of 87.5\% compared with 83.8\% for the best baseline, a consensus quality of Kendall's $W = 0.823$, and a clinical expert approval rating of 8.9/10, indicating that structured consensus modelling with a practical threshold of $W > 0.7$ can yield more accurate and stable recommendations while preserving guideline traceability and GRADE-based evidence appraisal. At the same time, the approach remains complementary to, rather than a replacement for, human expertise: potential biases in training data, dependence on current guidelines and the inherent complexity of ethically sensitive or rare cases underline the need for careful human oversight when deploying such systems in real-world clinical practice.

\section*{Acknowledgments}

The authors thank the medical experts who participated in the clinical evaluation and provided valuable feedback on system design and clinical applicability. We acknowledge the participating cancer centers for providing anonymized clinical cases for validation studies. This research was supported in part by grants from the National Science Foundation and the National Institutes of Health.

\bibliographystyle{IEEEtran}
\bibliography{arxiv}

\end{document}